\documentclass[pra,twocolumn,showpacs,preprintnumbers,amsmath,amssymb]{revtex4}

\usepackage{graphicx}
\usepackage{dcolumn}
\usepackage{bm}
\newcommand{\bra}[1]{\langle#1|}
\newcommand{\ave}[1]{\langle#1\rangle}

\newcommand{\ket}[1]{|#1\rangle}

\begin{document}

\title{Optimized experimental settings for the best detection of quantum nonlocality}

\author{Bruno Bellomo}
\author{Rosario Lo Franco}
\email{lofranco@fisica.unipa.it}
\author{Giuseppe Compagno}
\homepage{http://www.fisica.unipa.it/~compagno}
\affiliation{CNISM and Dipartimento di Scienze Fisiche ed Astronomiche,
Universit\`{a} di Palermo, via Archirafi 36, 90123 Palermo, Italy}

\date{\today}

\begin{abstract}
Nonlocality lies at the core of quantum mechanics from both a fundamental and applicative point of view. It is typically revealed by a Bell test, that is by violation of a Bell inequality, whose success depends both on the state of the system and on parameters linked to experimental settings. This leads to find, given the state, optimized parameters for a successful test. Here we provide, for a quite general class of quantum states, the explicit expressions of these optimized parameters and point out that, for a continuous change of the state, the corresponding suitable experimental settings may unexpectedly vary discontinuously. We finally show in a paradigmatic open quantum system that this abrupt ``jump'' of the experimental settings may even occur during the time evolution of the system. These jumps must be taken into account in order not to compromise the correct detection of nonlocality in the system.
\end{abstract}


\maketitle

Violation of a Bell inequality indicates, in two-qubit (spin-$1/2$-like) systems, the presence of quantum correlations nonreproducible by any classical local model (nonlocal correlations) \cite{bell,clauser}. A violation is identified by values of the Bell function $B$, consisting of a combination of correlations averages, larger than a maximum classical threshold  $B_\mathrm{class}$. For a given system the Bell function does depend both on the state of the system and on parameters linked to spin-like measurement directions that correspondingly fix the experimental settings. Many experimental tests have proven the occurrence of Bell inequality violations in different contexts as entangled-polarized photons \cite{kwiat2001Nature,groblacher2007Nature,zeilinger2009NatPhys}, photon-atom systems \cite{mohering2004PRL}, atom-atom systems \cite{wineland2001Nature} and recently also superconducting Josephson phase qubits \cite{martinis2009Nature}. Bell inequalities have also been extended to characterize the presence of nonlocality in multiqubit systems (Mermin inequalities) \cite{mermin1990PRL,horodecki2009RMP}. Nonlocal correlations also play a crucial role in quantum information science, as in the realization of device-independent and security-proof quantum key distribution protocols \cite{acin2006PRL,gisin2007natphoton}.

For pure states of any degree of entanglement it is possible to show that, by appropriately choosing the Bell parameters, violation of Bell inequality always occurs \cite{gisin1991PLA}. This is not the case for mixed states which may present entanglement without any corresponding direct violation of a Bell inequality \cite{werner1989PRA}. Indeed, mixed states take place as a consequence of unavoidable environmental noise which causes decoherence and a nonlocal correlations decay during the evolution \cite{bellomo2008bell,bellomo2009ASL}. In order to test at best the presence of nonlocality, it is essential to adopt optimized Bell parameters (OBPs) so that the maximum value of the Bell function $B_\mathrm{max}$ is measured. These OBPs in turn determine the appropriate experimental settings which allow to perform the optimal Bell test in order to state that, if $B_\mathrm{max}\leq B_\mathrm{class}$, quantum nonlocality is surely nondetectable directly. Because of the dependence of the OBPs on the state of the system, these are expected to change in time. If one wants to exploit nonlocal features of the system at time $t$, it is required to detect at best the presence of quantum nonlocal correlations at that time. Therefore, one needs to know how these OBPs vary in time so to correspondingly adjust the experimental settings and measure $B_\mathrm{max}(t)$. A procedure to obtain the OBPs and the corresponding maximum value of the Bell function for an arbitrary two-qubit state is known \cite{horodecki1995PLA}. However, the explicit expressions of these OBPs have never been reported before but for some very limited class of pure quantum states \cite{gisin1991PLA,nielsenchuang}.

Here we firstly give the explicit expressions of the optimized Bell parameters for a quite general class of pure and mixed quantum states showing moreover that, changing the quantum state in a continuous way, the OBPs may unexpectedly undergo a discontinuous change. These expressions are given for a two-spin-$1/2$ system where the parameters are the angles fixing directions in real space. Nevertheless, these parameters can be translated into the suitable experimental settings for any two-qubit system realization in different physical frameworks. We finally consider the inescapable environmental effects on the evolution of the two-qubit system and their influence on the Bell test. In a paradigmatic open quantum system we show that the abrupt discontinuous change of the OBPs may occur during the time evolution and this must be considered in order not to compromise the correct detection of nonlocality in the system.

\section{\label{Bell}Quantum nonlocality}
In the following we use the Clauser-Horne-Shimony-Holt (CHSH) form of Bell inequality \cite{clauser,clauser1974}, which is the most suitable for an experimental test of nonlocality in bipartite quantum systems composed by spin-$1/2$ objects. Let the operator $\mathcal{O}_S=\mathcal{O}_S(\theta_S,\phi_S)$ be a spin observable with eigenvalues $\pm1$ associated to the spin $S=1,2$ defined by $\mathcal{O}_S=\textbf{O}_S\cdot\bm{\sigma}_S$, where
$\textbf{O}_S\equiv(\sin\theta_S\cos\phi_S,\sin\theta_S\sin\phi_S,\cos\theta_S)$ is the unit vector indicating an arbitrary direction in the spin space and $\bm{\sigma}_S=(\sigma_x^S,\sigma_y^S,\sigma_z^S)$ is the Pauli matrices vector. The CHSH-Bell inequality associated to a two-spin-$1/2$ state $\hat{\rho}$ is then $B(\hat{\rho})\leq2$ \cite{clauser,clauser1974,horodecki1995PLA}, where $B(\hat{\rho})$ is the Bell function defined as
\begin{equation}\label{CHSHBellinequality}
B(\hat{\rho})=|\ave{\mathcal{O}_1\mathcal{O}_2}+\ave{\mathcal{O}_1\mathcal{O}'_2}
+\ave{\mathcal{O}'_1\mathcal{O}_2}-\ave{\mathcal{O}'_1\mathcal{O}'_2}|,
\end{equation}
where $\ave{\mathcal{O}_1\mathcal{O}_2}=\mathrm{Tr}\{\hat{\rho}\mathcal{O}_1\mathcal{O}_2\}$ is the correlation function of observables $\mathcal{O}_1$, $\mathcal{O}_2$ and $\mathcal{O}'_S\equiv\mathcal{O}_S(\theta'_S,\phi'_S)$. If, given the state $\hat{\rho}$, it is possible to find a set of angles $\{\theta_1,\theta'_1,\theta_2,\theta'_2\}$ and $\{\phi_1,\phi'_1,\phi_2,\phi'_2\}$ such that the CHSH-Bell inequality is violated ($B(\hat{\rho})>2$) the correlations are nonlocal and cannot be reproduced by any classical local model.

A necessary and sufficient condition for violating the CHSH-Bell inequality by an arbitrary two-spin-$1/2$ mixed state is known \cite{horodecki1995PLA}. Once introduced the $3\times3$ real matrix $T_\rho$, defined by the matrix elements $t_{mn}=\textrm{Tr}(\hat{\rho}\sigma_n\otimes\sigma_m)$, where the $\sigma$'s are the Pauli matrices, one must build the matrix $U_\rho\equiv T_\rho^\mathrm{T} T_\rho$, a symmetric matrix which can be therefore diagonalized ($T_\rho^\mathrm{T}$ indicates the transposed matrix of $T_\rho$). The maximum of the Bell function $B(\hat{\rho})$ results then to be given by $B_\mathrm{max}(\hat{\rho})=2\sqrt{\mathrm{max}_{j>k}\{u_j+u_k\}}$, where $j,k=1,2,3$ and
$u_j$'s are the three eigenvalues of the matrix $U_\rho$. A procedure to obtain the angles giving the maximum value of the Bell function is also provided \cite{horodecki1995PLA}.

\section{\label{Xstates}Maximum of Bell function}
In this section we give the maximum value of the Bell function for the class of states whose density matrix $\hat{\rho}_X$, in the standard computational basis $\mathcal{B}=\{\ket{1}\equiv\ket{11},\ket{2}\equiv\ket{10},\ket{3}\equiv\ket{01},\ket{4}\equiv\ket{00}\}$, has a X structure of the kind
\begin{equation}\label{Xstatesdensitymatrix}
   \hat{\rho}_X = \left(
\begin{array}{cccc}
  \rho_{11} & 0 & 0 & \rho_{14}  \\
  0 & \rho_{22} & \rho_{23} & 0 \\
  0 & \rho_{23}^* & \rho_{33} & 0 \\
  \rho_{14}^* & 0 & 0 & \rho_{44} \\
\end{array}
\right).
\end{equation}
This class of states is sufficiently general to include the two-qubit states most considered both theoretically and experimentally, like Bell states (pure two-qubit maximally entangled states) and Werner states (mixture of Bell states with white noise) \cite{nielsenchuang,bellomo2008PRA,horodecki2009RMP}. Moreover, a X structure density matrix arises in a wide variety of physical situations \cite{hagley1997PRL,bose2001,kwiat2001Nature,pratt2004,wang2006}. A further remarkable aspect of the X states is that, under various kinds of dynamics, the initial X structure is maintained during the evolution \cite{bellomo2007PRL,bellomo2008PRA}.

Using the criterion previously described to obtain $B_{\mathrm{max}}$, the three eigenvalues $u$'s, in terms of the density matrix elements, are found to be
\begin{eqnarray}\label{uXstate}
&u_1=4(|\rho_{14}|+|\rho_{23}|)^2,\quad
u_2=(\rho_{11}+\rho_{44}-\rho_{22}-\rho_{33})^2,&\nonumber\\
&u_3=4(|\rho_{14}|-|\rho_{23}|)^2,&
\end{eqnarray}
as also already reported \cite{derkacz2005PRA}. Being $u_1$ always larger than $u_3$, the maximum of Bell function for X states results to be
\begin{equation}
B_\mathrm{max}(\hat{\rho})=2\sqrt{u_1+\mathrm{max}_{j=2,3}\{u_j\}}.
\end{equation}

\section{\label{Expsettings} Optimized Bell parameters}
Here we give the OBPs, that is the angles $\theta$'s and $\phi$'s maximizing the Bell function $B(\hat{\rho})$ of Eq.~\eqref{CHSHBellinequality}, for the X states class of Eq.~\eqref{Xstatesdensitymatrix}. Following a known procedure \cite{horodecki1995PLA} we find the explicit OBPs, which result in two different sets of angles in correspondence of the two regions $u_2\geq u_3$ (region 1) and $u_3\geq u_2$ (region 2).

\textbf{Set 1: $u_2\geq u_3$.} The OBPs in this case are given by
\begin{eqnarray}\label{thetaphiangles1}
\theta_1&=&\pi/2,\ \theta'_1=0,\nonumber\\
\theta_2&=&\frac{\pi}{2}-\mathrm{sign}(\rho_{11}+\rho_{44}-\rho_{22}-\rho_{33})\arctan\sqrt{\frac{u_2}{u_1}},\nonumber\\ \theta'_2&=&\pi-\theta_2,\nonumber\\
\phi_1&=&-\left[\arg(\rho_{14})+\arg(\rho_{23})\right]/2,\quad\phi'_1=0,\nonumber\\
\phi_2&=&\phi'_2=\left[\arg(\rho_{23})-\arg(\rho_{14})\right]/2,
\end{eqnarray}
where the the function $\mathrm{sign}(x)$ is $+1$ if $x\geq0$ and $-1$ if $x<0$.

\textbf{Set 2: $u_3\geq u_2$.} The OBPs in this case are
\begin{eqnarray}\label{thetaphiangles2}
\theta_1&=&\theta'_1=\theta_2=\theta'_2=\pi/2,\nonumber\\
\phi_1&=&-\left[\arg(\rho_{14})+\arg(\rho_{23})\right]/2,\nonumber\\
\phi'_1&=&\phi_1+\mathrm{sign}(|\rho_{23}|-|\rho_{14}|)\pi/2,\nonumber\\
\phi_2&=&\frac{\arg(\rho_{23})-\arg(\rho_{14})}{2}+\arctan\sqrt{\frac{u_3}{u_1}},\nonumber\\
\phi'_2&=&\frac{\arg(\rho_{23})-\arg(\rho_{14})}{2}-\arctan\sqrt{\frac{u_3}{u_1}}.
\end{eqnarray}

Let us observe that the angles of the two sets are not equal when $u_2=u_3$. This means that, when this condition is satisfied, two different OBPs and so experimental settings do give the same value of $B_\mathrm{max}$. Moreover, considering the OBPs as function of the quantum state, to a slight continuous change of the quantum state crossing the point $u_2=u_3$, it corresponds an unexpected finite ``jump'' of the suitable experimental settings.

It is worth to emphasize the importance to have an analytical result for the two-qubit configuration here considered. In fact, if an analytical criterion to find OBPs was absent, numerical procedures would be necessary. This is the case, for example, in systems of more than two qubits where Mermin inequalities are considered to test nonlocality \cite{horodecki2009RMP}. On the basis on the two-qubit result, one may expect that a finite jump of the OBPs take place even for the multiqubit configuration. However, if this jump was present, starting from one of the OBPs sets, numerical procedures could give rise to serious problems in identifying the point of this possible OBPs jump because of the finite distance between the two sets of parameters.

It is clear that the angles of Eqs.~\eqref{thetaphiangles1} and \eqref{thetaphiangles2}, which determine the orientations of the measurement apparatus in real space giving $B_\mathrm{max}$ for two spin-$1/2$ objects, do correspond to different experimental settings for a general two-qubit system. In cavity QED, for example, qubits are represented by either two-level Rydberg atoms or cavities having one or zero photons; Ramsey zones followed by field ionization detectors are then the devices suitable for measuring the qubit along an arbitrary pseudospin direction, whose angles $\theta,\phi$ are controlled by the classical microwave field amplitude and the atom-field interaction time inside the Ramsey zone \cite{gerry1996PRA,lofranco2005PRA,lofranco2007PRA,haroche2006book}. In the context of traveling polarized photons, where the CHSH-Bell inequality was originally applied \cite{aspect1982PRL}, the spin levels are usually coded in the horizontal ($H$) and vertical ($V$) polarizations of the photon, the angles $\phi$'s are usually set at zero and the angles $\theta$'s are related to the angles $\theta_\textrm{p}$'s of photon polarizers by $\theta_\textrm{p}=\theta/2$ \citep{clauser,nielsenchuang}; the measurement is finally performed by photodetectors. In solid-state physics with Josephson phase qubits, coded by the two first levels of a well potential, pseudospin directions are selected by rotating the qubit states using microwave pulses before performing measurements by a pulse of current along the $z$ axis of the Bloch sphere (basis states of the qubit) \cite{martinis2009Nature}.

\section{\label{dynamicalexample} Optimized Bell parameters in a dynamical context}
So far we have found the optimized parameters (OBPs) that give the best condition for the Bell test on a given two-qubit X state. From an experimental point of view, one firstly needs to know the density matrix of the state on which to measure $B_\mathrm{max}$ and confirm if there is a violation of the CHSH-Bell inequality. This can be done by quantum state tomography \cite{kwiat2001Nature}. Therefore, in presence of a time-evolving quantum state, one may establish the form of the state with its density matrix elements at the time of the measurement. The experimental settings of the devices for the Bell test can be then appropriately adjusted according to our indications. In this section we give a dynamical example where the OBPs not only are crucial to reveal nonlocality, but they also present an unexpected behavior.

Consider a system of two identical independent qubits, namely 1 and 2, each embedded in a zero-temperature bosonic environment. Each qubit has ground and excited states $\ket{0}$, $\ket{1}$ separated by a transition frequency $\omega_0$. For our study we need the two-qubit reduced density matrix at the time $t$. This can be accomplished by exploiting a procedure based on the knowledge of the single-qubit dynamics \cite{bellomo2007PRL}. The dynamics of the single part ``qubit-environment'' is described, in the case of bosonic reservoir under rotating wave and dipole approximations, by the Hamiltonian \cite{petru}
\begin{equation}\label{Hamiltonian}
\hat{H}=\hbar \omega_0 \hat{\sigma}_+\hat{\sigma}_-+\sum_k \hbar\left[\omega_k \hat{b}_k^\dag \hat{b}_k+\left(g_k \hat{\sigma}_+\hat{b}_k+g_k^* \hat{\sigma}_-\hat{b}_k^\dag\right)\right],
\end{equation}
where $\sigma_+=\ket{1}\bra{0}$, $\sigma_-=\ket{0}\bra{1}$ are the qubit raising and lowering operators, $b_k^\dag$, $b_k$ are the photon creation and annihilation operators and $g_k$ is the coupling constant of the mode $k$ with frequency $\omega_k$. It is found that, for a zero-temperature environment, the single-qubit evolution under the Hamiltonian of Eq.~(\ref{Hamiltonian}) can be described in terms of a single function $q(t)$, whose square modulus $|q(t)|^2$ can be linked to the single-qubit excited state population. In the basis $\{\ket{1},\ket{0}\}$, the single-qubit reduced density matrix $\hat{\rho}^S(t)$ can be then written as \cite{petru}
\begin{equation}\label{roS}
\hat{\rho}^S(t)=\left(%
\begin{array}{cc}
\rho^S_{11}(0)|q(t)|^2  & \rho^S_{10}(0)q(t)\\\\
\rho^S_{01}(0)q^*(t)  & \rho^S_{00}(0)+ \rho^S_{11}(0)(1-|q(t)|^2) \\
\end{array}\right).
\end{equation}
The fact that the density matrix $\hat{\rho}^S(t)$ can be expressed in terms of a single function $q(t)$ is general, the information on the environment spectral density and coupling constants being contained in the explicit time dependence of the function $q(t)$ itself. The two-qubit density matrix $\hat{\rho}(t)$ is then obtained \cite{bellomo2007PRL}, its elements depending only on their initial values and on the function $q(t)$, regardless of the environment structure. Hereafter, for the sake of simplicity, we shall indicate $|q(t)|^2\equiv|q|^2$.

We choose as two-qubit initial state the extended Werner-like state defined as \cite{bellomo2008PRA}
\begin{equation}\label{EWLstate}
    \hat{\rho}^\Phi(0)=r\ket{\Phi}\bra{\Phi}+(1-r)I_4/4,
\end{equation}
where $r$ is the purity parameter, $I_4$ the $4\times4$ identity matrix and $\ket{\Phi}=\alpha\ket{01}+\beta e^{i\delta}\ket{10}$ is the one-excitation Bell-like state with $\alpha,\beta$ positive real and $\alpha^2+\beta^2=1$. $\hat{\rho}^\Phi(0)$ reduces to a Werner state for $\alpha=\beta=1/\sqrt{2}$ and $\delta=0,\pi$, that is when its pure part becomes a Bell state. For $r=0$ the state above becomes a totally mixed state, while for $r=1$ it reduces to the Bell-like (pure) state $\ket{\Phi}$. This state has an X form which is maintained during the dynamics due to the spin-boson Hamiltonian of Eq.~\eqref{Hamiltonian}. We can analyze the dependence of the dynamics of $B_\mathrm{max}$ on the initial mixedness and degree of entanglement.

\begin{figure}
\includegraphics[width=5.4 cm, height=4.5 cm]{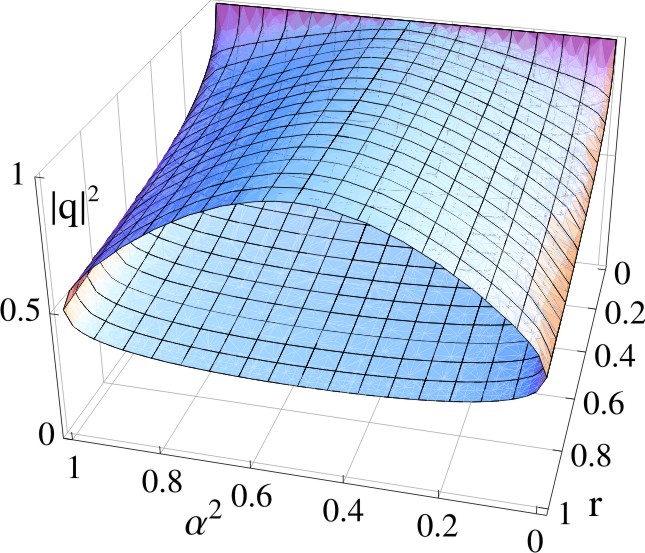}
\caption{\label{fig:qsurface}\footnotesize \textbf{Surface of the quantum state condition $u_2=u_3$ in the $(\alpha^2,r,|q|^2)$ space.} Given $\alpha^2, r$ two values of $|q|^2$ ($|q_1|^2$ and $|q_2|^2$) satisfy $u_2=u_3$. The points outside the surface represent the region 1 where $u_2>u_3$, while those inside the surface represent the region 2 where $u_3>u_2$.}
\end{figure}

Starting from the two-qubit state defined above, the expressions of $u_1$, $u_2$ and $u_3$ of Eq.~\eqref{uXstate} are found to be in terms of $|q|^2$ as
\begin{equation}\label{uEWL}
u_1=u_3=4\alpha^2\beta^2r^2|q|^4,\ u_2=[1-2|q|^2+(1-r)|q|^4]^2.
\end{equation}
The condition $u_2=u_3$ determines a closed surface in the 3D space ($\alpha^2,r,|q|^2$) given in Fig.~\ref{fig:qsurface}, with two values $|q_1|^2$, $|q_2|^2$ for a fixed couple $\alpha^2, r$.
The points enclosed by this surface represent the region 2 ($u_3>u_2$), while the points outside represent the region 1 ($u_2>u_3$). Dynamically $|q|^2$ changes with time and Fig.~\ref{fig:qsurface} clearly shows that, given $\alpha^2$ and $r$, when $|q|^2$ crosses the surface at time $t$, the OBPs immediately before and after this time are given respectively by the two different sets 1 or 2. This in turn gives place to a discontinuous change with time of the experimental settings giving the maximum of Bell function.

\begin{figure}
\includegraphics[width=8 cm, height=8 cm]{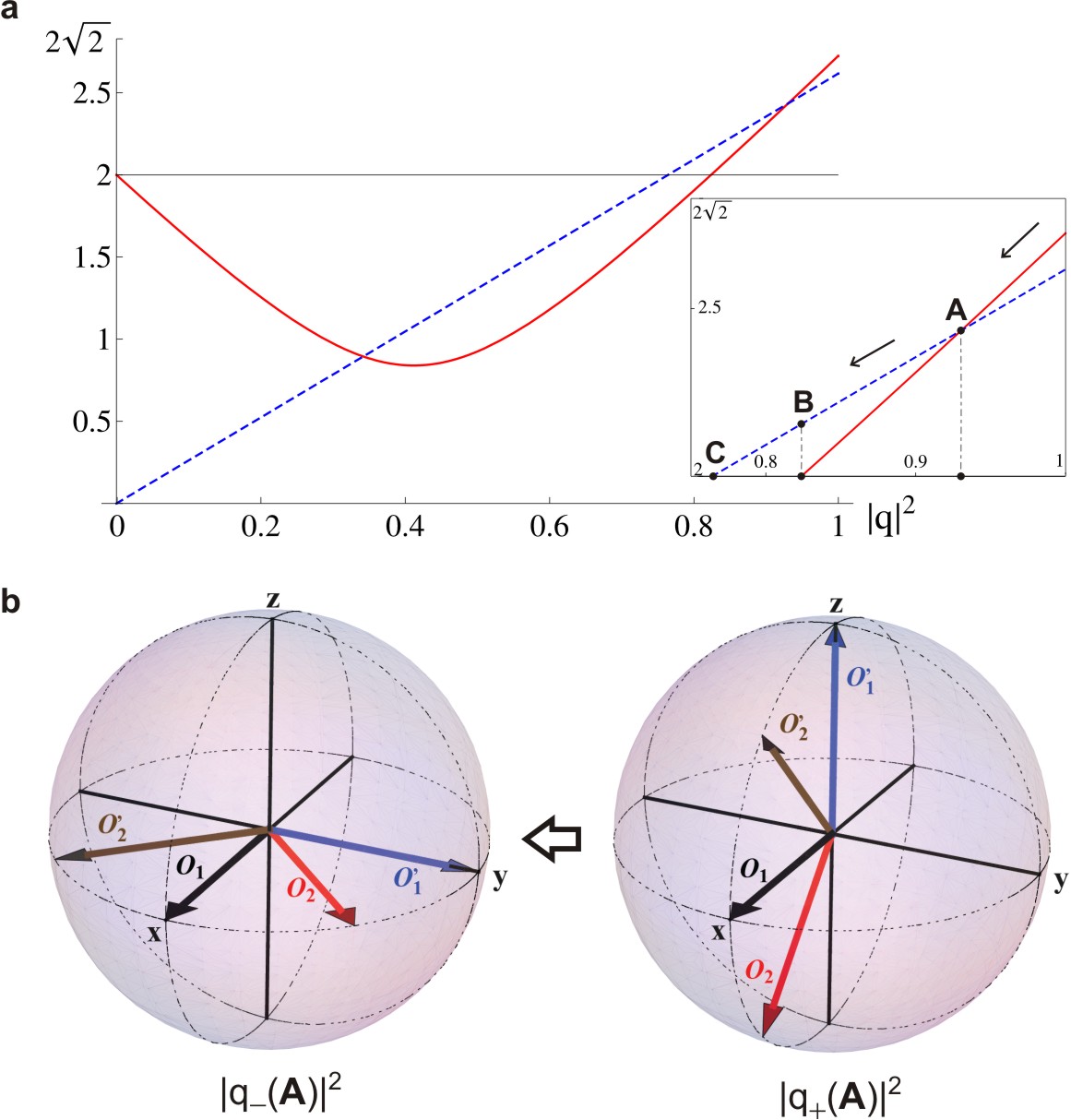}
\caption{\label{fig:B1B2measdirections}\footnotesize \textbf{Dynamical behavior of $B_1$ and $B_2$ with abrupt ``jump'' of the OBPs.} \textbf{a}, Plot of $B_1=2\sqrt{u_1+u_2}$ (solid curve) and $B_2=2\sqrt{u_1+u_3}$ (dashed curve) as a function of $|q|^2$, for $\alpha^2=0.3$ and $r=1$. The inset evidences the violation region where $B_1,B_2>2$. The value $|q(\textbf{A})|^2$ indicates the change from branch 1 to branch 2 (going from right to left), while between the values $|q(\textbf{B})|^2$ and $|q(\textbf{C})|^2$ only $B_2\equiv B_\mathrm{max}$ is larger than two. \textbf{b}, The spin-$1/2$ measurement vectors determined by the OBPs are illustrated (for $\delta=0$) in correspondence of $|q(\textbf{A})|^2$. The abrupt jump of the OBPs is well evidenced while passing from $|q_+(\textbf{A})|^2$ to $|q_-(\textbf{A})|^2$. Immediately before and after the crossing point $|q(\textbf{A})|^2$, the spin measurement vectors lie on the $x$-$z$ plane and $x$-$y$ plane, respectively.}
\end{figure}
In order to examine this surprising aspect, we plot in Fig.~\ref{fig:B1B2measdirections} both $B_1=2\sqrt{u_1+u_2}$ and $B_2=2\sqrt{u_1+u_3}$ as a function of $|q|^2$ for $\alpha^2=0.3$ and $r=1$. In this case the initial state of Eq.~\eqref{EWLstate} is a pure non-maximally entangled state and $B_\mathrm{max}=\mathrm{max}\{B_1,B_2\}$.
In the plot one moves from right to left with time ($|q|^2$ going from 1 to 0) and $B_\mathrm{max}$ is given by $B_1$ (branch 1; solid curve) before $|q(\textbf{A})|^2$ and by $B_2$ (branch 2; dashed curve) after it. The region of violation of CHSH-Bell inequality ($B_\mathrm{max}>2$) is zoomed in the inset. At value $|q(\textbf{A})|^2$ there is the passage from branch 1 to branch 2 and the corresponding abrupt discontinuous ``jump'' of the OBPs (see part \textbf{b} of Fig.~\ref{fig:B1B2measdirections}) which pass from set 1 to set 2. Moreover, for values $|q(\textbf{C})|^2\leq |q|^2\leq  |q(\textbf{B})|^2$, using the parameters of set 1 (giving the maximal violation when $|q|^2\geq|q(\textbf{A})|^2$) one finds no violation of Bell inequality while using those of set 2 one finds a violation.

The environment determines the explicit time evolution of $|q|^2\equiv|q(t)|^2$. For Markovian environments $|q(t)|^2$ decays exponentially \cite{petru} and the values $|q(\textbf{A})|^2$ and $|q(\textbf{B})|^2$ are typically reached after a short time. Differently for structured environments inhibiting the population decay, such as photonic crystals, $|q(t)|^2$ decay can be slow \cite{bellomo2008trapping} and the crossing value $|q(\textbf{A})|^2$ could be reached after a rather long time. Another case is that of non-Markovian environments where collapses and revivals of $|q(t)|^2$ may be present, such as in high-$Q$ cavities \cite{bellomo2007PRL}, and the surface of Fig.~\ref{fig:qsurface} could be crossed several times during the evolution.

Therefore, when one needs to detect the presence of nonlocal correlations without ambiguity, the striking behavior of the OBPs here shown highlights the importance of our analysis to fix the most appropriate experimental settings to reveal nonlocality.

\section{\label{Conclusions}Conclusions}
In this paper we have found the optimized parameters determining the experimental settings suitable to reveal at best the presence of nonlocality in a bipartite spin-$1/2$-like system. We have reported the explicit dependence on the state of these optimized Bell parameters (OBPs) for a general class of quantum states. We have found that this class of states can be divided in two parts each of them corresponding to a different set of OBPs. We have also found that even if the state of the system changes continuously, as when it evolves with time, a surprising abrupt jump of these OBPs may occur in proximity of some states. In particular, we have shown that these jumps of the experimental settings that maximize the Bell function occur during the evolution of a paradigmatic two-qubit open system. This phenomenon must therefore be taken into account in any experimental test aiming at obtaining the best confirmation of the presence of nonlocality in a quantum system.

\end{document}